\documentclass[review]{elsarticle}

\usepackage{bm}
\usepackage{graphicx}
\usepackage[utf8]{inputenc}
\usepackage[english]{babel}

\begin{document}

\title{Nambu mechanics for stochastic magnetization dynamics}

\author[cea]{Pascal Thibaudeau\corref{cor1}}
\ead{pascal.thibaudeau@cea.fr}
\address[cea]{CEA DAM/Le Ripault, BP 16, F-37260, Monts, FRANCE}

\author[cea,lmpt]{Thomas Nussle}
\ead{thomas.nussle@cea.fr}
\address[lmpt]{CNRS-Laboratoire de Mathématiques et Physique Théorique (UMR 7350), Fédération de Recherche "Denis Poisson" (FR2964), Département de Physique, Université de Tours, Parc de Grandmont, F-37200, Tours, FRANCE}

\author[lmpt]{Stam Nicolis}
\ead{stam.nicolis@lmpt.univ-tours.fr}

\cortext[cor1]{Corresponding author}

\begin{abstract}
The Landau-Lifshitz-Gilbert (LLG) equation describes the dynamics of a damped magnetization vector
that can be understood as a generalization of Larmor spin precession. The LLG equation cannot be deduced from the Hamiltonian framework, by introducing a coupling to a usual bath, but requires the introduction of additional constraints. It is shown that these constraints can be formulated elegantly and consistently in the framework of dissipative Nambu mechanics. This has many consequences for both the variational principle  and for  topological aspects of hidden symmetries that control conserved quantities. We particularly study how  the damping terms  of dissipative Nambu mechanics affect  the consistent interaction of magnetic  systems with  stochastic reservoirs and derive a master equation for the magnetization.  The  proposals  are supported by numerical studies using symplectic integrators that preserve the topological structure of Nambu equations. These results are compared to computations performed by direct  sampling of the stochastic equations and by using closure assumptions for the moment equations, deduced from the master equation. 
\end{abstract}


\maketitle

\section{Introduction}\label{introduction}
In micromagnetism, the transverse Landau-Lifshitz-Gilbert (LLG) equation 
\begin{equation}
(1+\alpha^2)\frac{\partial s_i}{\partial t}=\epsilon_{ijk}\omega_j({\bm s})s_k+\alpha(\omega_i({\bm s})s_js_j-\omega_j({\bm s})s_js_i)\label{LLG}
\end{equation}
describes the dynamics of a magnetization vector ${\bm s}\equiv{\bm M}/M_s$ with $M_s$ the saturation magnetization. This equation can be seen as a generalization of Larmor spin precession,  for a collection of elementary classical magnets evolving in an effective pulsation ${\bm\omega}=-\frac{1}{\hbar}\frac{\delta H}{\delta{\bm s}}=\gamma{\bm B}$ and within a magnetic medium, characterized by a damping constant $\alpha$ and a gyromagnetic ratio $\gamma$ \cite{bertotti_nonlinear_2009}. $H$ is here identified as a scalar  functional of the magnetization vector and can be consistently generalized to include spatial derivatives of the magnetization vector \cite{aharoni_introduction_2000} as well. Spin-transfer torques, that are, nowadays, of particular practical relevance~\cite{miltat_spin_2001,berkov_spin-torque_2008} can be, also, taken into account in this formalism. In the following, we shall work in units where $\hbar=1$, to simplify notation. 

It is well known that this equation cannot be derived from a Hamiltonian variational principle, with the damping effects  described by coupling the magnetization to a bath, by deforming the Poisson bracket of Hamiltonian mechanics, even though the Landau--Lifshitz equation itself is Hamiltonian. The reason is that the damping cannot be described by a ``scalar'' potential, but by a ``vector'' potential. 

This has been made manifest \cite{holyst_dissipative_1992} first by an analysis of the quantum version of the Landau-Lifshitz equation for damped spin motion including arbitrary spin length, magnetic anisotropy and many interacting quantum spins. In particular, this analysis has revealed that the damped spin equation of motion is an example of metriplectic dynamical system \cite{turski_dissipative_1996}, an approach which tries to unite symplectic, nondissipative and metric, dissipative dynamics into one common mathematical framework. This dissipative system has been seen afterwards nothing but a natural combination of semimetric dynamics for the dissipative part and Poisson dynamics for the conservative ones \cite{nguyen_dirac_2001}. As a consequence, this provided a canonical description for any constrained dissipative systems through an extension of the concept of Dirac brackets developed originally for conservative constrained Hamiltonian dynamics. Then, this has culminated recently by observing the underlying geometrical nature of these brackets as certain $n$-ary generalizations of Lie algebras, commonly encountered in conservative Hamiltonian dynamics \cite{de_azcarraga_n_2010}. However, despite the evident progresses obtained, no clear direction emerges for the case of dissipative $n$-ary generalizations, and even no variational principle have been formulated, to date, that incorporates such properties.            

What we shall show in this paper is that it is, however, possible to describe the Landau--Lifshitz--Gilbert equation by using the variational principle of Nambu mechanics and to describe the damping effects as the result of introducing dissipation by suitably deforming the Nambu--instead of the Poisson--bracket. In this way we shall find, as a bonus, that it is possible to deduce  the relation between longitudinal and transverse damping of the magnetization, when writing the appropriate master equation for the probability density. To achieve this in a Hamiltonian formalism requires additional assumptions, whose provenance can, thus, be understood as the result of the properties of Nambu mechanics. We focus here on the essential points; a fuller account will be provided in future work. 

Neglecting damping effects, if one sets $H_1\equiv -\bm{\omega}\cdot\bm{s}$ and $H_2\equiv \bm{s}\cdot\bm{s}/2$, eq.(\ref{LLG}) can be recast in the form
\begin{equation}
\label{Nambu}
\frac{\partial s_i}{\partial t}=\left\{s_i,H_1,H_2\right\},
\end{equation}
where for any functions $A$, $B$, $C$ of ${\bm s}$,
\begin{equation}
\label{NambuBracket}
\left\{A,B,C\right\}\equiv\epsilon_{ijk}\frac{\partial A}{\partial s_i}\frac{\partial B}{\partial s_j}\frac{\partial C}{\partial s_k}
\end{equation} 
is the Nambu-Poisson (NP) bracket, or Nambu bracket, or Nambu triple bracket, a skew-symmetric object, obeying both the Leibniz rule and the Fundamental Identity \cite{nambu_generalized_1973,ibanez_dynamics_1997}. One can see immediately that both $H_1$ and $H_2$ are constants of motion, because of the anti-symmetric property of the bracket.
This provides the generalization of Hamiltonian mechanics to phase spaces of arbitrary dimension; in particular it does not need to be even. This is a way of taking into account constraints and provides a natural framework for describing the magnetization dynamics, since the magnetization vector has, in general, three components.  

The constraints--and the symmetries--can be made manifest, by noting that it is possible to express vectors and vector fields in, at least, two ways, that can be understood as special cases of Hodge decomposition. 

For the three--dimensional case that is of interest here, this means that a vector field ${\bm V}({\bm s})$ can be expressed in the ``Helmholtz representation''~\cite{marsden_introduction_1999} in the following way 
\begin{equation}
\label{Helmholtz}
V_i\equiv\epsilon_{ijk}\frac{\partial A_k}{\partial s_j}+\frac{\partial \Phi}{\partial s_i}
\end{equation}
where ${\bm A}$ is a vector potential and $\Phi$ a scalar potential.

On the other hand, this same vector field  ${\bm V}(\bm{s})$ can be decomposed according to the ``Monge representation'' \cite{griffiths_principles_2014}
\begin{equation}
\label{Monge}
V_i\equiv\frac{\partial C_1}{\partial s_i}+C_2\frac{\partial C_3}{\partial s_i}
\end{equation}
which defines  the ``Clebsch-Monge potentials'', $C_i$.

If one identifies as the Clebsch--Monge potentials, $C_2\equiv H_1$, $C_3\equiv H_2$ and $C_1\equiv D$,
\begin{equation}
V_i=\frac{\partial D}{\partial s_i}+H_1\frac{\partial H_2}{\partial s_i},
\end{equation}
and the vector field $\bm{V}(\bm{s})\equiv\dot{\bm{s}}$, then one immediately finds that eq.~(\ref{Nambu}) takes the form
\begin{equation}
\label{Nambudissipative}
\frac{\partial\bm{s}}{\partial t} = \left\{\bm{s},H_1,H_2\right\} + \nabla_{\bm{s}} D
\end{equation}
that identifies the contribution of the dissipation in this context, as the expected generalization from usual Hamiltonian mechanics. In the absence of the Gilbert term, dissipation is absent. 

More generally, the evolution equation for any function, $F({\bm s})$ can be written as~\cite{axenides_strange_2010} 
\begin{equation}
\frac{\partial F}{\partial t}=\left\{F,H_1,H_2\right\}+\frac{\partial D}{\partial s_i}\frac{\partial F}{\partial s_i}\label{dissipativenambu}
\end{equation}
 for a dissipation function $D(\bm{s})$. 
 
The equivalence between the Helmholtz and the Monge representation implies the existence of freedom of redefinition for the potentials, $C_i$ and $D$ and $A_i$ and $\Phi$. This freedom expresses the symmetry under symplectic transformations, that can be interpreted as diffeomorphism transformations, that leave the volume invariant. These have consequences for the equations of motion.

For instance, the dissipation described by the Gilbert term in the Landau--Lifshitz--Gilbert equation~(\ref{LLG})
\begin{equation}
\frac{\partial D}{\partial s_i} \equiv \alpha({\tilde\omega}_i({\bm s})s_js_j-{\tilde\omega}_j({\bm s})s_js_i)
\end{equation}
cannot be derived from a scalar potential, since the RHS of this expression is not curl--free, so the function $D$ on the LHS is not single valued; but it does conserve the norm of the magnetization, i.e. $H_2$. Because of the Gilbert expression, both $\bm{\omega}$ and ${\bm\eta}$ are rescaled such as $\tilde{\bm \omega}\equiv{\bm \omega}/(1+\alpha^2)$ and ${\bm\eta}\rightarrow {\bm\eta}/(1+\alpha^2)$. So there are two questions: (a) Whether it can lead to stochastic effects, that can be described in terms of deterministic chaos and/or (b) Whether its effects can be described by a bath of ``vector potential'' excitations. The first case was described, in outline in ref.~\cite{tranchida_quantum_2015}, where the role of an external torque was shown to be instrumental; the second will be discussed in detail in the following sections. While, in both cases, a stochastic description, in terms of a probability density on the space of states is the main tool, it is much easier to present for the case of a bath, than for the case of deterministic chaos, which is much more subtle.  

Therefore, we shall now couple our magnetic moment to a bath of fluctuating degrees of freedom, that will be described by a stochastic process. 

\section{Nambu dynamics in a macroscopic bath}

To this end, one couples linearly the deterministic system such as (\ref{dissipativenambu}), to a stochastic process, i.e. a noise vector, random in time, labelled $\eta_i(t)$, whose law of probability is given. This leads to a 
system of stochastic differential equations, that can be written in the Langevin form 
\begin{equation}
\frac{\partial s_i}{\partial t}=\left\{s_i,H_1,H_2\right\}+\frac{\partial D}{\partial s_i}+e_{ij}({\bm s})\eta_j(t)
\label{randomnambu}
\end{equation}
where $e_{ij}(\bm{s})$ can be interpreted as the vielbein on the manifold, defined by the dynamical variables, $\bm{s}$. It should be noted that it is the vector nature of the dynamical variables that implies that the vielbein,
 must, also, carry indices. 
 
We may note that the additional noise term can be used to ``renormalize'' the precession frequency and, thus, mix, non-trivially, with the Gilbert term. This means that, in the presence of either, the other cannot be excluded. 
 
When this vielbein is the identity matrix, $e_{ij}(\bm{s})=\delta_{ij}$, the stochastic coupling to the noise is additive, whereas it is multiplicative otherwise. In that case, if the norm of the spin vector has to remain constant in time, then the gradient of $H_2$ must be orthogonal to the gradient of $D$ and $e_{ij}({\bm s})s_i=0$ $\forall j$.  
 
However, it is important to realize that, while the Gilbert dissipation term is not a gradient, the noise term, described by the vielbein is not so constrained. For additive noise, indeed, it is a gradient, while for the case of multiplicative noise studied by Brown and successors there can be  an interesting interference between the two terms, that is worth studying in more detail, within Nambu mechanics, to understand, better, what are the coordinate artifacts and what are the intrinsic features thereof.  

Because $\left\{{\bm s}(t)\right\}$, defined by  the eq.(\ref{randomnambu}), becomes  a stochastic  process,  we can define an instantaneous conditional probability distribution $P_\eta({\bm s},t)$, that depends, on the noise configuration and, also, on  the magnetization ${\bm s}_0$ at  the initial time  and which satisfies a continuity equation  in  configuration space
\begin{equation}
\frac{\partial P_\eta({\bm s},t)}{\partial t}+\frac{\partial \left(\dot{s}_iP_{\eta}({\bm s},t))\right)}{\partial s_i}=0.
\end{equation}
An equation for $\langle P_\eta \rangle$ can be formed, which becomes an average over all the possible realizations of the noise, namely
\begin{equation}
\frac{\partial \langle P_\eta \rangle}{\partial t}+\frac{\partial \langle\dot{s}_iP_\eta\rangle}{\partial s_i}=0,
\label{averageP}
\end{equation}
once the distribution law of $\left\{{\bm\eta}(t)\right\}$ is provided. It is important to stress here that this implies that the backreaction of the spin degrees of freedom on the bath can be neglected--which is by no means obvious. One way to check this is by showing that no ``runaway solutions'' appear. This, however, does not exhaust all possibilities, that can be found by working with the Langevin equation directly. For non--trivial vielbeine, however, this is quite involved, so it is useful to have an approximate solution in hand. 

To be specific, we consider a noise, described by the Ornstein-Uhlenbeck process \cite{uhlenbeck_theory_1930} of intensity $\Delta$ and autocorrelation  time $\tau$, 
\begin{eqnarray}
\langle\eta_i(t)\rangle&=&0\nonumber\\
\langle\eta_i(t)\eta_j(t')\rangle&=&\frac{\Delta}{\tau}\delta_{ij}e^{-\frac{|t-t'|}{\tau}}\nonumber
\end{eqnarray}
where the higher point  correlation functions are deduced from Wick's theorem and 
which can be shown  to become a white noise process, when $\tau\rightarrow 0$. We assume  that the solution to eq.(\ref{averageP}) converges, in the sense of average over-the-noise, to an equilibrium distribution, that is normalizable and, whose correlation functions, also, exist. While this is, of course, not at all obvious to prove, evidence can be found by numerical studies, using  stochastic integration methods that  preserve the symplectic structure of the Landau--Lifshitz  equation, even under perturbations~(cf. \cite{thibaudeau_thermostatting_2012} for earlier work). 
\subsection{Additive noise}
Walton \cite{walton_rate_1986} was one of the first to consider the introduction of an additive noise into an LLG equation and remarked that it may lead to a Fokker-Planck equation, without entering into details. To see this more thoroughly and to illustrate our strategy, we consider the case of additive noise, i.e. when $e_{ij}=\delta_{ij}$ in our framework. By including eq.(\ref{randomnambu}) in (\ref{averageP}) and in the limit of white noise, expressions like $\langle\eta_iP_\eta\rangle$ must be defined  and can be evaluated by either an expansion of the Shapiro-Loginov formulae of differentiation \cite{shapiro_formulae_1978} and taking the limit of $\tau\rightarrow 0$, or, directly, by applying the  Furutsu-Novikov-Donsker theorem \cite{furutsu_statistical_1963,novikov_functionals_1964,klyatskin_stochastic_2005}. This leads to
\begin{equation}
\langle \eta_iP_\eta\rangle=-{\tilde\Delta}\frac{\partial \langle P_\eta\rangle}{\partial s_i}.\label{additivenoise}
\end{equation}
where $\tilde\Delta\equiv\Delta/(1+\alpha^2)$. Using the dampened current vector $J_i\equiv \{s_i,H_1,H_2\}+\frac{\partial D}{\partial s_i}$, the (averaged) probability density $\langle P_\eta\rangle$ satisfies the following equation
\begin{equation}
\frac{\partial \langle P_\eta\rangle}{\partial t}+\frac{\partial}{\partial s_i}\!\left(J_i \langle P_\eta\rangle \right)-{\tilde{\tilde\Delta}}\frac{\partial^2 \langle P_\eta\rangle}{\partial s_i \partial s_i}=0
\end{equation}
where ${\tilde{\tilde\Delta}}\equiv\Delta/(1+\alpha^2)^2$ and which is of the Fokker-Planck form \cite{risken_fokker-planck_1989}. This last partial differential equation can be solved directly by several numerical methods, including a finite-element computer code or can lead to ordinary  differential equations for the moments of ${\bm s}$.

For example, for the average of the magnetization, one obtains the evolution equation 
\begin{equation}
\frac{d \langle s_i\rangle}{dt}=-\int d{\bm s}\,s_i\frac{\partial \langle P_\eta({\bm s},t)\rangle}{\partial t}=\left\langle J_i\right\rangle.
\end{equation}
For the case of Landau-Lifshitz-Gilbert in a uniform precession field ${\bm B}$,  we obtain the following equations, for  the first and second moments,
\begin{eqnarray}
\frac{d }{dt}\langle s_i \rangle&=&\epsilon_{ijk}{\tilde\omega}_j\langle s_k \rangle+\alpha [{\tilde\omega}_i\langle s_js_j\rangle-{\tilde\omega}_j\langle s_js_i\rangle]\label{dllba}\\
\frac{d}{dt}\langle s_is_j \rangle&=&{\tilde\omega}_l\left(\epsilon_{ilk}\langle s_ks_j\rangle+\epsilon_{jlk}\langle s_ks_i\rangle\right)+\alpha\left[{\tilde\omega}_i\langle s_ls_ls_j\rangle\right.\nonumber\\
&+&\left.{\tilde\omega}_j\langle s_ls_ls_i\rangle-2{\tilde\omega}_l\langle s_ls_is_j\rangle\right]+2{\tilde{\tilde\Delta}}\delta_{ij}
\label{dllbb}\label{secondOrderMomentsEvolution}
\end{eqnarray}
where ${\tilde{\bm \omega}}\equiv \gamma{\bm B}/(1+\alpha^2)$. In order to close consistently these equations, one can truncate the hierarchy of moments; either on the second $\langle\!\langle s_is_j\rangle\!\rangle=0$ or third cumulants $\langle\!\langle s_is_js_k\rangle\!\rangle=0$, i.e. 
\begin{eqnarray}
\langle s_i s_j\rangle&=&\langle s_i\rangle\langle s_j\rangle, \\
\langle s_is_js_k\rangle&=&\langle s_is_j\rangle \langle s_k\rangle +\langle s_is_k\rangle \langle s_j\rangle+\langle s_js_k\rangle \langle s_i\rangle\nonumber\\
&-&2\langle s_i\rangle \langle s_j\rangle \langle s_k\rangle.
\end{eqnarray}
Because the closure of the hierarchy is related to an expansion in powers of $\Delta$, for practical purposes, the validity of eqs.(\ref{dllba},\ref{dllbb}) is limited to low values of the coupling to the bath (that describes the fluctuations). For example, if one sets $\langle\!\langle s_is_j\rangle\!\rangle=0$, eq.(\ref{dllba}) produces an average spin motion independent of value that $\Delta$ may take. This is in contradiction with the numerical experiments performed by the stochastic integration and noise average of eq.(\ref{randomnambu}) quoted in reference \cite{tranchida_closing_2016} and by experiments. This means that it is mandatory to keep at least eqs.(\ref{dllba}) and (\ref{dllbb}) together in the numerical evaluation of the thermal behavior of the dynamics of the average thermal magnetization $\langle{\bm s}\rangle$. This was previously observed \cite{garanin_fokker-planck_1997,ma_langevin_2011} and circumvented by alternate second-order closure relationships, but is not supported by direct numerical experiments.

This can be illustrated by the following figure (\ref{Fig1}). 
\begin{figure}[thp]
\centering
\resizebox{0.99\columnwidth}{!}{\includegraphics{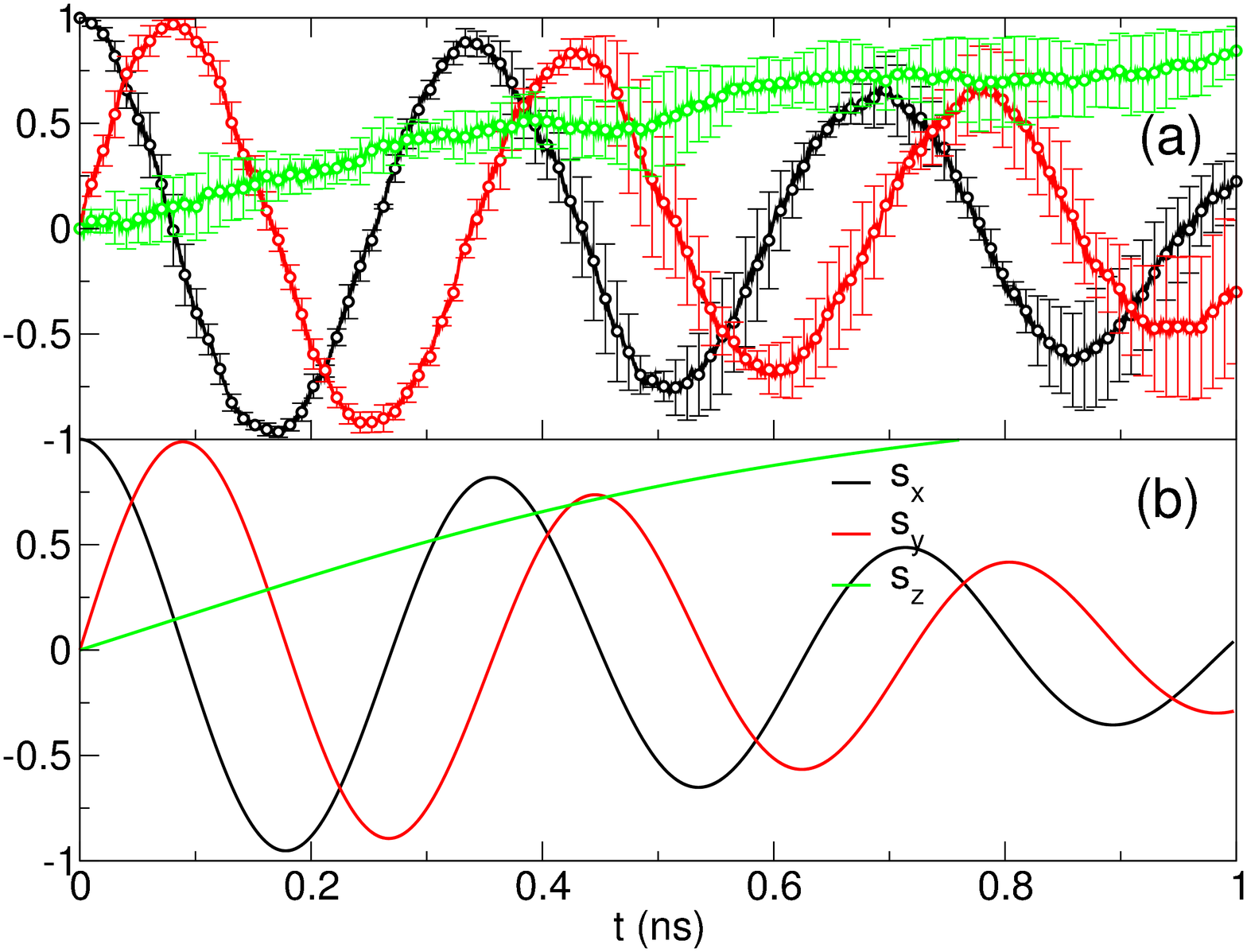}}
\caption{
Magnetization dynamics of a paramagnetic spin in a constant magnetic field, connected to an additive noise. The upper graphs (a) plot some of the first--order moments of the averaged magnetization vector over $10^2$ realizations of the noise, when the lower graphs (b) plot the associated model closed to the third-order cumulant (eqs.(\ref{dllba})-(\ref{dllbb}), see text). Parameters of the simulations~: \{$\Delta=0.13$~rad.GHz; $\alpha=0.1$; ${\bm\omega}=(0,0,18)$~rad.GHz; timestep $\Delta t=10^{-4}$ ns\}. Initial conditions: ${\bm s}(0)=\left(1,0,0 \right)$, $\langle s_i(0) s_j(0) \rangle=0$ but $\langle s_1(0)s_1(0) \rangle=1$. \label{Fig1}}
\end{figure}
For this given set of parameters, the agreement between the stochastic average and the effective model is fairly decent. As expected, for a single noise realization, the norm of the spin vector in an additive stochastic noise cannot be conserved during the dynamics, but, by the average-over-the-noise accumulation process, this is observed for very low values of $\Delta$ and very short times. However, this agreement with the effective equations is lost, when the temperature increases, because of the perturbative nature of the equations (\ref{dllba}-\ref{dllbb}). Agreement can, however, be  restored by imposing  this constraint in the effective equations, for a given order in perturbation of $\Delta$, by appropriate modifications of the hierarchical closing relationships $\langle\!\langle s_is_j\rangle\!\rangle=B_{ij}(\Delta)$ or $\langle\!\langle s_is_js_k\rangle\!\rangle=C_{ijk}(\Delta)$. 

\begin{figure}[thp]
\centering
\resizebox{0.99\columnwidth}{!}{\includegraphics{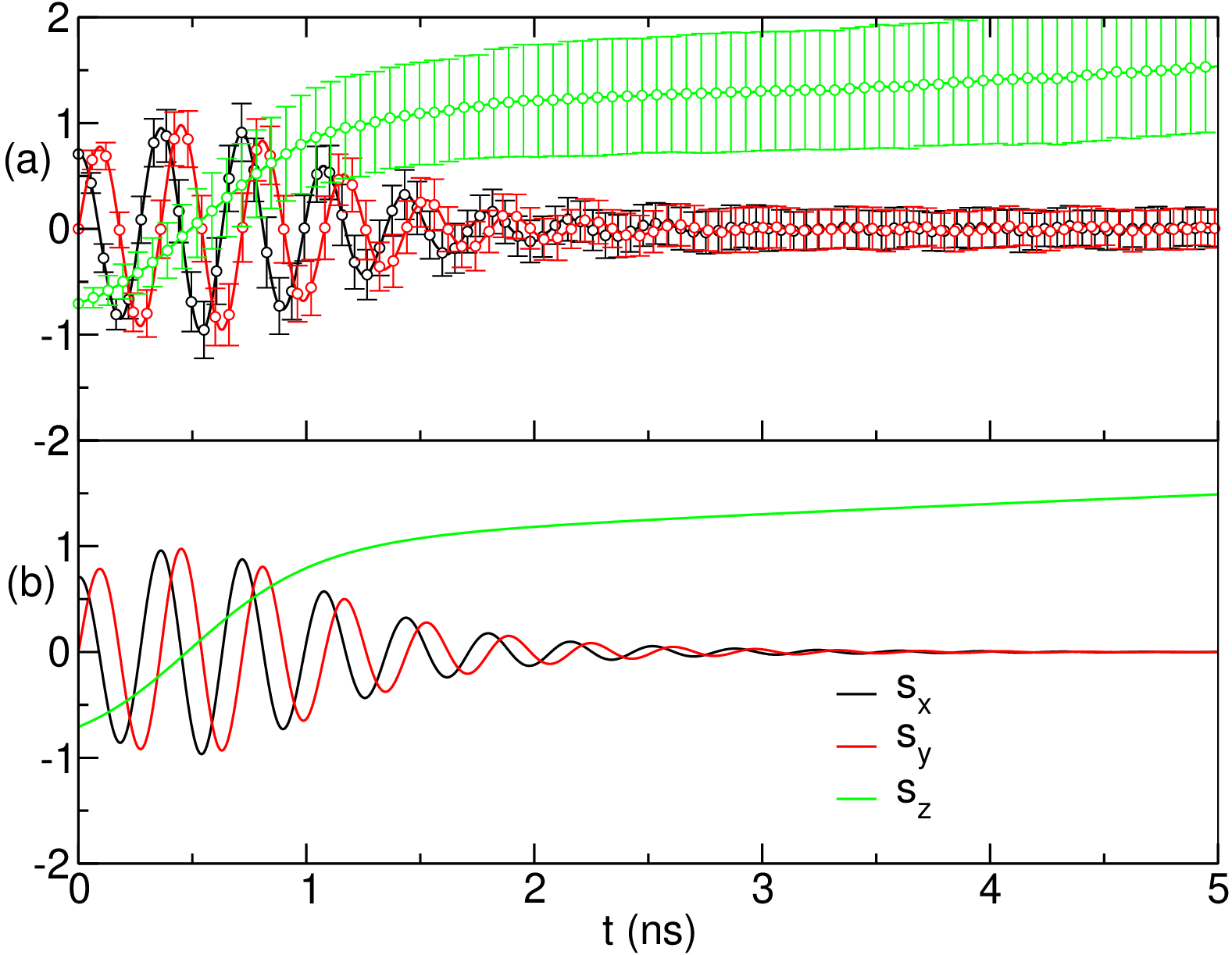}}
\caption{Magnetization dynamics of a paramagnetic spin in a constant magnetic field, connected to an additive noise. The upper graphs (a) plot some of the first--order moments of the averaged magnetization vector over $10^3$ realizations of the noise, when the lower graphs (b) plot the associated model closed to the third-order cumulant (eqs.(\ref{dllba})-(\ref{dllbb}), see text). Parameters of the simulations~: \{$\Delta=0.0655$~rad.GHz; $\alpha=0.1$; ${\bm\omega}=(0,0,18)$~rad.GHz; timestep $\Delta t=10^{-4}$ ns, ${\bm s}(0)=\langle{\bm s}(0)\rangle=(1/\sqrt{2},0,-1/\sqrt{2})$, $\langle s_is_j\rangle(0)=0$ except for (11)=1/2, (13)=(31)=-1/2, (33)=1/2 \}. \label{Fig2}}
\end{figure}

It is of some interest to study the effects of the choice of initial conditions. In particular, how the relaxation to equilibrium is affected by choosing a component of the initial magnetization along the precession axis in the effective model, e.g. 
$\bm{s}(0)=(1/\sqrt{2},0,-1/\sqrt{2})$ 
and by taking all the initial correlations,
\begin{equation}
\langle s_i(0)s_j(0)\rangle=\left(
\begin{array}{ccc}
\frac{1}{2}&0&-\frac{1}{2}\\
0&0&0\\
-\frac{1}{2}&0&\frac{1}{2}
\end{array}
\right)
\end{equation} 
The results are shown in figure (\ref{Fig2}).

\begin{figure}[thp]
\centering
\resizebox{0.99\columnwidth}{!}{\includegraphics{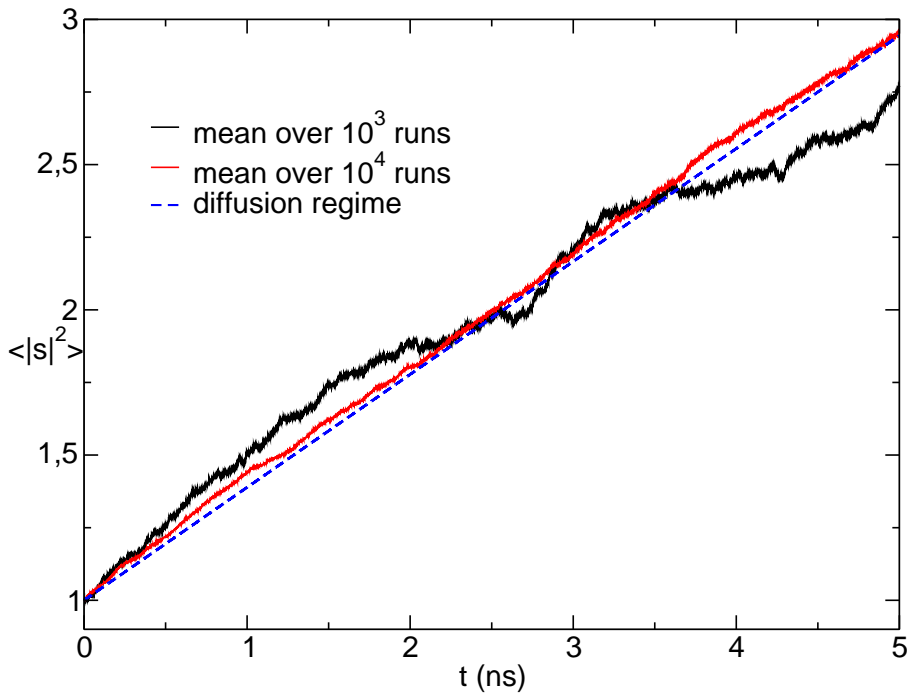}}
\caption{Mean square norm of the spin in the additive white noise case for the following conditions: integration step of $10^{-4}$ns; $\Delta=0.0655$ rad.GHz; ${\bm s}(0)=\left(0,1,0 \right)$; $\alpha=0.1$; ${\bm \omega}=(0,0,18)$~rad.GHz compared to the expected diffusion regime (see text).\label{Fig3}}
\end{figure}

Both in figures (\ref{Fig1}) and (\ref{Fig2}), it is observed that the average norm of the spin vector increases over time. This can be understood with the above arguments. In general, according to eq.(\ref{randomnambu}) and because ${\bm J}$ is a transverse vector,
\begin{equation}
(1+\alpha^2)s_i\frac{ds_i}{dt}=e_{ij}(\bm{s})s_i\eta_j(t).
\end{equation}
This equation describes how the LHS depends on the noise realization; so the average over the noise can be found by computing the averages of the RHS. The simplest case is that of the additive vielbein, $e_{ij}(\bm{s})=\delta_{ij}$.  Assuming that the average-over-the noise procedure and the time derivative commute, we have
\begin{equation}
\frac{d}{dt}\left\langle{\bm s}^2\right\rangle=\frac{2\langle s_i\eta_i\rangle}{1+\alpha^2}.
\end{equation}
For any Gaussian stochastic process, the Furutsu-Novikov-Donsker theorem states that
\begin{equation}
\langle s_i(t)\eta_i(t)\rangle=\int_{-\infty}^{+\infty}dt'\langle \eta_i(t)\eta_j(t')\rangle\left\langle\frac{\delta s_i(t)}{\delta \eta_j(t')}\right\rangle.\label{averageadditivemix}
\end{equation}
In the most general situation, the functional derivatives $\displaystyle{\frac{\delta s_i(t)}{\delta \eta_j(t')}}$ can be calculated \cite{tranchida_functional_2016}, and eq.(\ref{averageadditivemix}) admits simplifications in the white noise limit. In this limit, the integration is straightforward and we have 
\begin{equation}
\left\langle\bm{s}^2(t)\right\rangle={\bm s}^2(0)+6{\tilde{\tilde\Delta}}t,
\end{equation}
which is a conventional diffusion regime. It is also worth noticing that when computing the trace of (\ref{secondOrderMomentsEvolution}), the only term which remains is indeed 
\begin{equation}
\frac{d}{dt}\langle s_is_i\rangle=6\tilde{\tilde {\Delta}}
\end{equation}
which allows our effective model to reproduce exactly the diffusion regime. Figure (\ref{Fig3}) compares the time evolution of the average of the square norm spin vector. 
Numerical stochastic integration of eq.(\ref{randomnambu}) is tested by increasing the size of the noise sampling and reveals a convergence to the predicted linear diffusion regime.

\subsection{Multiplicative noise}
Brown \cite{brown_thermal_1963} was one of the first to propose a non--trivial vielbein, that takes the form $e_{ij}(\bm{s})=\epsilon_{ijk}s_k/(1+\alpha^2)$ for the LLG equation. We notice, first of all, that it is present, even if $\alpha=0$, i.e. in the absence of the Gilbert term. Also,  that,
 since the determinant of this matrix $[e]$ is zero, this vielbein is not invertible. Because of its natural transverse character, this vielbein preserves the norm of the spin for any realization of the noise, once a dissipation function $D$ is chosen, that has this property. In the white-noise limit, the average over-the-noise continuity equation (\ref{averageP}) cannot be transformed strictly to a Fokker-Planck form. This time
\begin{equation}
\langle \eta_iP_\eta\rangle=-{\tilde\Delta}\frac{\partial}{\partial s_j}\left(e_{ji}\langle P_\eta\rangle\right),
\end{equation}
which is a generalization of the additive situation shown in eq.(\ref{additivenoise}). The continuity equation thus becomes
\begin{equation}
\frac{\partial \langle P_\eta\rangle}{\partial t}+\frac{\partial}{\partial s_i}\!\left(J_i \langle P_\eta\rangle \right)-{\tilde{\tilde\Delta}}\frac{\partial}{\partial s_i}\left(e_{ij}\frac{\partial}{\partial s_k}\left(e_{kj}\langle P_\eta\rangle\right)\right)=0.
\label{FPmult}
\end{equation}
What deserves closer attention is, whether, in fact, this equation is invariant under diffeomeorphisms of the manifold~\cite{zinn-justin_quantum_2011} defined by the vielbein, or whether it breaks it to a subgroup thereof. This will be presented in future work.
In the context of magnetic thermal fluctuations, this continuity equation was encountered several times in the literature \cite{risken_fokker-planck_1989,garcia-palacios_langevin-dynamics_1998}, but obtaining it from first principles is more cumbersome than our latter derivation, a remark already quoted \cite{shapiro_formulae_1978}. Moreover, our derivation presents the advantage  of being  easily generalizable to  non-Markovian noise distributions \cite{tranchida_closing_2016,thibaudeau_non-markovian_2016,tranchida_colored-noise_2016}, by simply keeping the partial derivative equation on the noise with the continuity equation, and solving them together.

Consequently, the evolution equation for the average magnetization is now supplemented by a term provided by a non constant vielbein and one has
\begin{equation}
\label{multaveragemag}
\frac{d \langle s_i\rangle}{dt}=\left\langle J_i\right\rangle+{\tilde{\tilde\Delta}}\left\langle\frac{\partial e_{il}}{\partial s_k}e_{kl}\right\rangle.
\end{equation}
With the vielbein proposed by Brown and assuming  a constant external field, one gets
\begin{eqnarray}
\frac{ d\langle s_i\rangle}{dt}&=&\epsilon_{ijk}\tilde\omega_j\langle s_k\rangle+\alpha\left(\tilde\omega_i\langle s_js_j\rangle-\tilde\omega_j\langle s_js_i\rangle\right)\nonumber\\
&&-\frac{2\Delta}{(1+\alpha^2)^2}\langle s_i\rangle.\label{muldllba}
\end{eqnarray} 
This equation highlights both a transverse part, coming from the average over the probability current ${\bm J}$ and a longitudinal part, coming from the average over the extra vielbein term. By imposing, further, the second-order cumulant approximation $\langle\!\langle s_is_j\rangle\!\rangle=0$, i.e. ``small''  fluctuations to keep the distribution of ${\bm s}$ gaussian, a single equation can be obtained, in which a longitudinal relaxation time $\tau_L\equiv (1+\alpha^2)^2/2\Delta$ may be identified.

\begin{figure}[thp]
\centering
\resizebox{0.99\columnwidth}{!}{\includegraphics{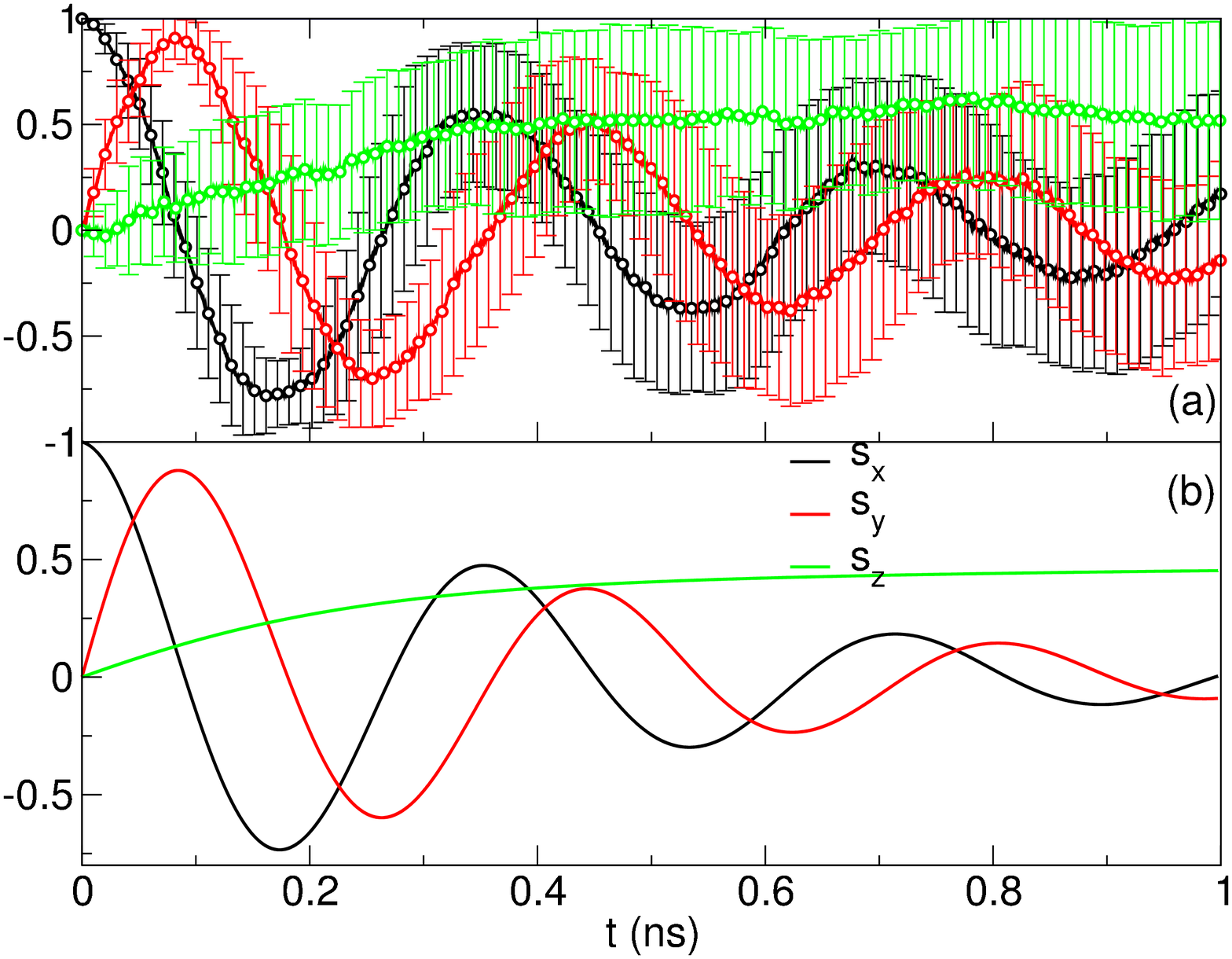}}
\caption{
Magnetization dynamics of a paramagnetic spin in a constant magnetic field, connected to a multiplicative noise. The upper graphs (a) plot some of the first--order moments of the averaged magnetization vector over $10^2$ realizations of the noise, when the lower graphs (b) plot the associated model closed to the third-order cumulant (eq.(\ref{muldllba}), see text). Parameters of the simulations~: \{$\Delta=0.65$~rad.GHz; $\alpha=0.1$; ${\bm\omega}=(0,0,18)$~rad.GHz; timestep $\Delta t=10^{-4}$ ns\}. Initial conditions: ${\bm s}(0)=\left(1,0,0 \right)$, $\langle s_i(0) s_j(0) \rangle=0$ but $\langle s_x(0)s_x(0) \rangle=1$. \label{Fig4}}
\end{figure}

This is illustrated by the content of figure (\ref{Fig4}). 
In that case, the approximation $\langle\!\langle s_is_js_j\rangle\!\rangle=0$ has been retained in order to keep two sets of equations, three for the average magnetization components and nine on the average second-order moments, that have been solved simultaneously using an eight-order Runge-Kutta algorithm with variable time-steps. This is the same numerical implementation that has been followed for the studies of the additive noise, solving eqs.(\ref{dllba}) and (\ref{dllbb}) simultaneously. We have observed numerically that, as expected, the average second-order moments are symmetrical by an exchange of their component indices, both for the multiplicative and the additive noise. Interestingly, by keeping identical the number of random events taken to evaluate the average of the stochastic magnetization dynamics between the additive and multiplicative noise, we observe a greater variance in the multiplicative case.

\begin{figure}[thp]
 \centering
 \resizebox{0.99\columnwidth}{!}{\includegraphics{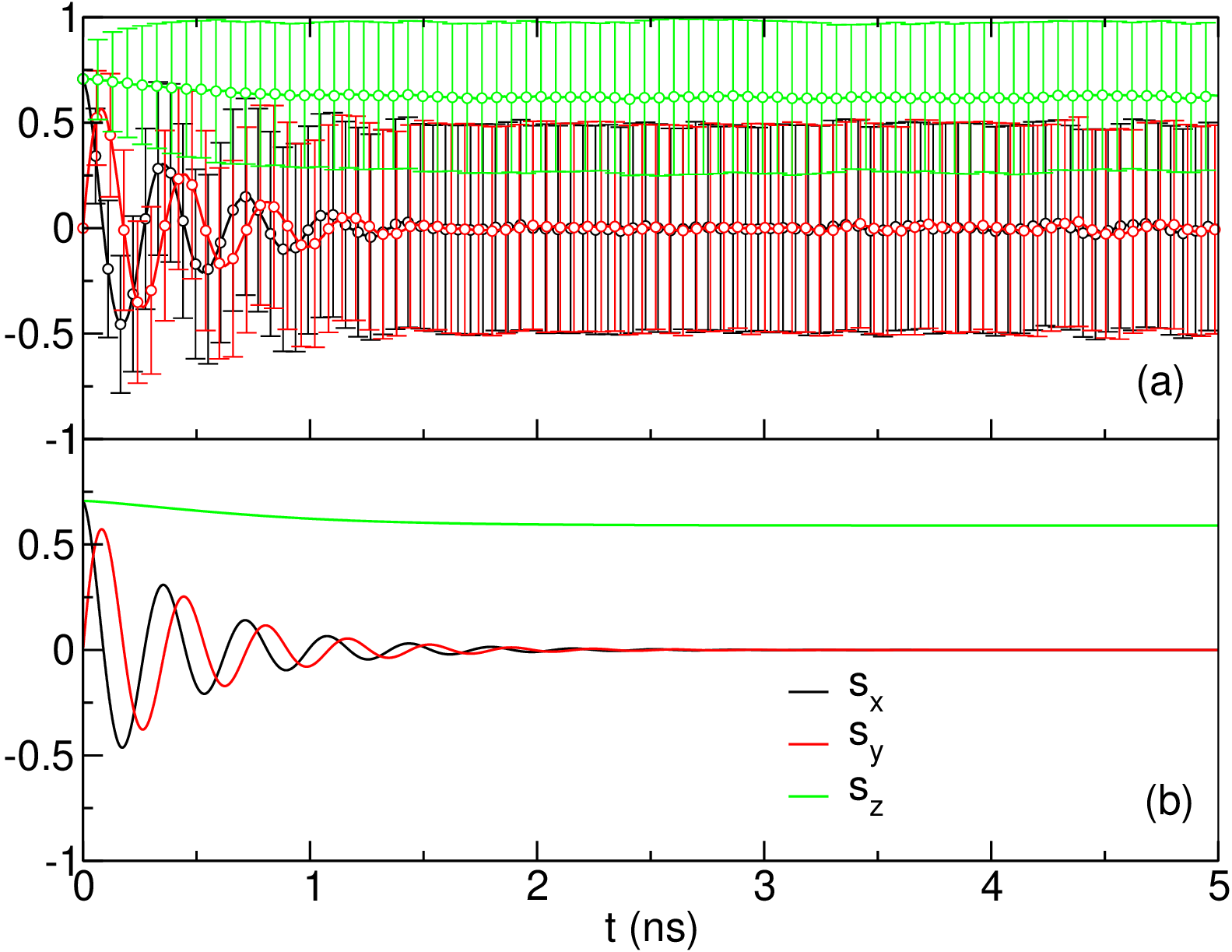}}
 \caption{Magnetization dynamics of a paramagnetic spin in a constant magnetic field, connected to a multiplicative noise. The upper graphs (a) plot some of the first--order moments of the averaged magnetization vector over $10^4$ realizations of the noise, when the lower graphs (b) plot the associated model closed to the third-order cumulant (eq.(\ref{muldllba}), see text). Parameters of the simulations~: \{$\Delta=0.65$~rad.GHz; $\alpha=0.1$; ${\bm\omega}=(0,0,18)$~rad.GHz; timestep $\Delta t=10^{-4}$ ns\}. Initial conditions: ${\bm s}(0)=\left(1/\sqrt{2},0,1/\sqrt{2} \right)$, $\langle s_i(0) s_j(0) \rangle=0$ except for $\langle s_1(0)s_1(0) \rangle=\langle s_1(0)s_3(0) \rangle=\langle s_3(0)s_3(0) \rangle=1/2$. \label{Fig5}}
\end{figure}

As we have done in the additive noise case, we will also investigate briefly the behavior of this equation under different initial conditions, and in particular with a non vanishing component along the z-axis. This is illustrated by the content of figure (\ref{Fig5}). 
It is observed that for both figures (\ref{Fig4}) and (\ref{Fig5}), the average spin converges to the same final equilibrium state, which depends ultimately on the value of the noise amplitude, as shown by equation (\ref{FPmult}). 

\section{Discussion}
Magnetic systems describe vector  degrees of freedom, whose Hamiltonian dynamics implies constraints. These constraints can be naturally taken into account within Nambu mechanics, that generalizes Hamiltonian mechanics to phase spaces of odd number of dimensions. In this framework, dissipation can be described by gradients  that are not single--valued and thus do not define scalar baths, but vector baths, that, when coupled to external torques, can lead to chaotic dynamics. The vector baths can, also, describe non-trivial geometries and, in that case, as we have shown by direct numerical study, the stochastic description leads to a coupling between longitudinal and transverse relaxation. This can be, intuitively, understood within Nambu mechanics, in the following way: 

The dynamics consists in rendering one of the Hamiltonians, $H_1\equiv\bm{\omega}\cdot\bm{s}$, stochastic, since $\bm{\omega}$ becomes a stochastic process, as it is sensitive to the noise terms--whether these are described by Gilbert dissipation or coupling to an external bath.  Through the Nambu equations, this dependence is ``transferred'' to $H_2\equiv||\bm{s}||^2/2$. This is one way of realizing the insights the Nambu approach provides. 

In practice, we may summarize our numerical results as follows: 
 
When the amplitude of the noise is small, in the context of Langevin-dynamics formalism for linear systems and for the numerical modeling of {\it small} thermal fluctuations in micromagnetic systems, as for a linearized stochastic LLG equation, the rigorous method of Lyberatos, Berkov and Chantrell might be thought to apply~\cite{lyberatos_method_1993} and be expected to be equivalent to the approach presented here. Because this method expresses the approach to equilibrium of every moment, separately, however,  it is restricted to the limit of small fluctuations around an equilibrium state and, as expected, cannot capture the transient regime of average magnetization dynamics, even for low temperature. This is a useful check.

We have also  investigated the behaviour of this system under different sets of initial conditions as it is well-known and has been thoroughly studied in \cite{bertotti_nonlinear_2009} that in the multiplicative noise case (where the norm is constant) this system can show strong sensitivity to initial conditions and it is possible, using stereographic coordinates to represent the dynamics of this system in 2D. In our additive noise case however, as the norm of the spin is not conserved, it is not easy to get long run behavior of our system and in particular equilibrium solutions. Moreover as we no longer have only two independent components of spin, it is not possible to obtain a 2D representation of our system and makes it more complicated to study maps displaying limit cycles, attractors and so on. Thus understanding the dynamics under different initial conditions would require something more and, as it is beyond the scope of this work, will be done elsewhere.

Therefore, we have focused on studying the effects of the presence of an initial longitudinal component and of additional, diagonal, correlations. No differences have been observed so far.

Another issue, that deserves further study, is how the probability density of the initial conditions is affected by the stochastic evolution. In the present study we have taken the initial probability density to be a $\delta-$function; so it will be  of interest to study the evolution of other initial distributions in detail, in particular, whether the averaging procedures commute--or not. In general, we expect that they won't. This will be reported in future work. 

Finally, our study can be readily generalized since any vielbein can be expressed in terms of a diagonal, symmetrical and anti-symmetrical matrices, whose elements are functions of the dynamical variable ${\bm s}$. Because ${\dot{\bm s}}$ is a pseudovector (and  we do not consider that this additional property is acquired by the noise vector), this suggests that the anti-symmetric part of the vielbein should be the ``dominant'' one. Interestingly, by numerical investigations, it appears that there are no effects, that might depend on the choice of the noise connection for the stochastic vortex dynamics in two-dimensional easy-plane ferromagnets \cite{kamppeter_stochastic_1999}, even if it is known that for Hamiltonian dynamics, multiplicative and additive noises usually modify the dynamics quite differently, a point that also deserves further study.        
\bibliographystyle{unsrt}
\bibliography{cea-lmpt}
\end{document}